\documentclass[preprint,aps,amsmath,amssymb]{revtex4}

\newcommand {\bc}{\begin{center}}
\newcommand {\ec}{\end{center}}
\newcommand {\bea}{\begin{eqnarray}}
\newcommand {\eea}{\end{eqnarray}}
\newcommand {\be}{\begin{equation}}
\newcommand {\ee}{\end{equation}}

\def\lsim{\mathrel{\rlap{\lower4pt\hbox{\hskip1pt$\sim$}}
    \raise1pt\hbox{$<$}}}               
\def\gsim{\mathrel{\rlap{\lower4pt\hbox{\hskip1pt$\sim$}}
    \raise1pt\hbox{$>$}}}

\usepackage{graphicx}

\begin{document}


\title{Bulk viscosity and conformal symmetry breaking
in the dilute Fermi gas near unitarity}

\author{Kevin Dusling and Thomas~Sch\"afer}

\affiliation{Department of Physics, North Carolina State University,
Raleigh, NC 27695}

\begin{abstract}
The dilute Fermi gas at unitarity is scale invariant and its
bulk viscosity vanishes. We compute the leading contribution 
to the bulk viscosity when the scattering length is not 
infinite. A measure of scale breaking is provided by the 
ratio $(P-\frac{2}{3}{\cal E})/P$, where $P$ is the pressure
and ${\cal E}$ is the energy density. In the high temperature 
limit this ratio scales as $\frac{z\lambda}{a}$, where $z$ 
is the fugacity, $\lambda$ is the thermal wave length, and 
$a$ is the scattering length. We show that the bulk viscosity
$\zeta$ scales as the second power of this parameter, $\zeta
\sim (\frac{z\lambda}{a})^2 \lambda^{-3}$. 
\end{abstract}

\maketitle

\section{Introduction}
\label{sec_intro}

 The dilute Fermi gas at unitarity is a beautiful example of a 
scale and conformally invariant many body system. Scale invariance 
implies that thermodynamic properties of the system only depend on 
the dimensionless variable $n\lambda^3$, where $n$ is the density 
and $\lambda=[2\pi\hbar/(mT)]^{1/2}$ is the thermal de Broglie wave 
length. In the high temperature limit $n\lambda^3\ll 1$ and the gas 
is weakly interacting despite the fact that the two-body scattering 
length $a$ is tuned to infinity. In the low temperature regime $n
\lambda^2\leq  1$  the gas is strongly correlated. It was observed 
that in this limit the unitary Fermi gas is a very good liquid, 
characterized by a very small shear viscosity $\eta\lsim \hbar n$ 
\cite{OHara:2002,Schafer:2007pr,Kinast:2005zz}. Nearly perfect 
fluidity was also observed in the quark gluon plasma produced in 
heavy collisions at RHIC and the LHC, and it arises naturally in 
the context of holographic dualities 
\cite{Kovtun:2004de,Schafer:2009dj,Adams:2012th}.

 Scale invariance is broken if the Fermi gas is detuned from unitarity 
and the two-body scattering length is not infinite. A measure of 
scale invariance breaking is the difference $P-\frac{2}{3}{\cal E}$,
where $P$ is the pressure and ${\cal E}$ is the energy density. 
Tan showed that \cite{Tan:2005,Tan:2008}
\be 
P-\frac{2}{3}{\cal E} = \frac{\hbar^2{\cal C}}{12\pi ma}\,
\ee
where ${\cal C}$ is the contact density. At unitarity and in the high 
temperature limit ${\cal C}=4\pi\hbar n^2\lambda^2$ \cite{Yu:2009}. This 
implies that $(P-\frac{2}{3}{\cal E})/P\sim (n\lambda^3)(\lambda/a)$.
In the present work we address the question how broken scale invariance 
manifests itself in transport properties. The natural quantity to 
consider is the bulk viscosity $\zeta$ which vanishes in a scale invariant 
fluid \cite{Son:2005tj,Enss:2010qh,Castin:2011}. We will show that in 
the high temperature limit $\zeta$ scales as the shear viscosity times 
the square of the conformal breaking parameter $(n\lambda^3)(\lambda/a)$.
An analogous relation was derived by Weinberg in the case of a 
relativistic gas \cite{Weinberg:1971mx}. He showed that $\zeta \sim 
\eta (c_s^2-c^2/3)^2$, where $c_s$ is the speed of sound and $c$ is the
speed of light. In a scale invariant relativistic fluid $P={\cal E}/3$
and $c_s^2=c^2/3$. 

 The physical mechanism for generating bulk viscosity in a 
non-relativistic gas of structureless particles is subtle. In a typical 
non-relativistic gas, such as air, bulk viscosity arises from rotational 
and vibrational excitations of the air molecules \cite{WangChang:1964}. 
In equilibrium, if the gas is compressed or expanded internal energy is 
transferred from center of mass motion to internal degrees of freedom. 
This transfer requires scattering processes, and if  these reactions 
are slow then the system will fall out of equilibrium. The departure of 
the pressure from its equilibrium value is related to bulk viscosity. 
In polyatomic gases bulk viscosity also arises from energy transfer 
between different species, or from chemical non-equilibration. In 
systems in which the number of particles is not conserved, such as 
a gas of phonons, bulk viscosity may arise from number changing 
processes. None of these mechanisms operates in a dilute Fermi 
gas above the superfluid transition. 

 In a relativistic gas bulk viscosity arises from non-zero particles 
masses, often combined with number changing processes. This is the case, 
for example, in a dilute gases of quarks and gluons \cite{Arnold:2006fz}, 
or a dilute gas of pions \cite{Lu:2011df}. In a nearly scale invariant 
gas, such as the quark gluon plasma, masses only arise from interactions 
and the effective mass is of the form $m\sim gT$, where $g$ is the QCD 
coupling constant. In this case bulk viscosity is governed by the scale 
breaking part of the effective mass, $\tilde{m}^2=(1-T^2\frac{\partial}
{\partial T^2})m^2$ \cite{Jeon:1995zm,Arnold:2006fz}. In QCD, scale breaking 
arises from the logarithmic running of $g$ with the temperature $T$. We 
will show that a similar mechanism operates in the dilute Fermi gas 
detuned from unitarity. Bulk viscosity arises from the scale breaking 
part of a temperature and density dependent effective mass. The new 
ingredient compared to a relativistic plasma is that the momentum
dependence of the effective mass is also crucial. 

 This paper is organized as follows. In Sect.~\ref{sec_diag} we 
introduce a diagrammatic approach to the thermodynamic and single 
particle properties of the dilute Fermi gas. In Sect.~\ref{sec_qp_boltz}
we match this approach to a quasi-particle Boltzmann equation. In
Sect.~\ref{sec_CE} we solve the Boltzmann equation using the 
Chapman-Enskog procedure and determine bulk viscosity. Thermodynamic
relations can be found in the appendix. Note that in the remainder 
of this paper we will set $\hbar=k_B=1$. 

 Our work is related to a number of recent studies that address transport 
properties of nearly conformal non-relativistic fluids. Sum rules for 
the bulk viscosity were derived in \cite{Taylor:2010ju} and further 
elaborated in \cite{Enss:2010qh,Goldberger:2011hh,Hofmann:2011qs}.
The superfluid phase is characterized by three bulk viscosity 
coefficients \cite{Khalatnikov:1965}. Son showed that two of these 
have to vanish at unitarity \cite{Son:2005tj}. A calculation of 
$\zeta_{1,2,3}$ near unitarity based on a kinetic theory of phonons 
can be found in \cite{Escobedo:2009bh}. Finally, it is interesting to 
consider two-dimensional fluids. In two dimensions scale invariance 
is always broken by quantum mechanical effects. It was nevertheless 
observed experimentally that there is an almost undamped breathing 
mode at twice the trap frequency \cite{Vogt:2011np}. This experiment 
was recently studied in \cite{Taylor:2012pe}.

\section{High temperature expansion}
\label{sec_diag}
 
 The effective lagrangian for non-relativistic spin 1/2 fermions
interacting via a short range $s$-wave potential is 
\be 
\label{l_4f}
{\cal L} = \psi^\dagger \left( i\partial_0 +
 \frac{\nabla^2}{2m} \right) \psi 
 - \frac{C_0}{2} \left(\psi^\dagger \psi\right)^2 ,
\ee
where the coupling constant $C_0$ is determined by the $s$-wave
scattering length $a$. The precise relation depends on the 
regularization scheme. In dimensional regularization we find
$C_0 = 4\pi a/m$. The effective lagrangian can be partially 
bosonized using the Hubbard-Stratonovich transformation. 
Introducing an integral over a complex bosonic field $\Phi$
and shifting integration variables we can write
\be
\label{l_4f_hs}
{\cal L} = \psi^\dagger \left( i\partial_0 +
 \frac{\nabla^2}{2m} \right) \psi  
 + \left[ (\psi\sigma_+\psi)\Phi + {\it h.c.}\right]
+\frac{1}{C_0}|\Phi|^2\, . 
\ee
The integration over the fermion fields can now be carried
out and we find an effective action for the bosonic field $\Phi$,
\be
\label{s_ng_eff}
 S= -{\rm Tr}\left[\log\left(G^{-1}\left[\Phi,\Phi^*\right]\right)\right]+
     \int d^4x\, \frac{1}{C_0}|\Phi|^2,
\ee
where $G^{-1}$ is a $2\times 2$ matrix. In momentum space we have
\be
\label{ng_prop}
 G^{-1}\left[\Phi,\Phi^*\right] = 
 \left(\begin{array}{cc}
     p_0-\epsilon_p  & \Phi^* \\
     \Phi & p_0+\epsilon_p
 \end{array}\right),
\ee
with $\epsilon_p=p^2/(2m)$.

\subsection{Thermodynamic properties}

 We compute the thermodynamic potential $\Omega$ using the Matsubara 
formalism. We introduce a chemical potential for $\psi$, continue the 
fields to imaginary time $\tau$, and impose periodic/anti-periodic boundary 
conditions on the bosonic/fermionic fields. We evaluate the partition 
function by expanding the logarithm to quadratic order in $\Phi$, and 
then compute the Gaussian integral over $\Phi$. At high temperature 
higher order terms in $\Phi$ can be treated perturbatively. We find
\be 
\label{Omega_1l}
\Omega = T\sum_n\int \frac{d^3q}{(2\pi)^3}
   \log\left[ \chi(i\omega_n,q)\right]\, , 
\ee
where $\omega_n=2\pi n T$ are bosonic Matsubara frequencies
and $\chi(\omega_n,q)$ is the one loop particle-particle 
polarization function
\be 
\label{chi_pp}
 \chi(i\omega_n,q) = \int \frac{d^3k}{(2\pi)^3} 
    \Bigg\{ \frac{1-f_k-f_{k+q}}{i\omega_n -\xi_k-\xi_{k+q}}
       + \frac{1}{2\epsilon_k} \Bigg\} 
       - \frac{m}{4\pi a}\, .
\ee
Here, $\xi_k=\epsilon_k-\mu$ and $f_k=[\exp(\beta\xi_k)+1]^{-1}$ is 
the Fermi-Dirac distribution function. In deriving equ.~(\ref{chi_pp}) 
we have used the vacuum relation between $C_0$ and the scattering 
length. The Matsubara sum in equ.~(\ref{Omega_1l}) can be performed
using contour integration. The final result can be written in terms 
of the discontinuity of $\chi(\omega,q)$ along the real axis. We 
find 
\be 
\label{Omega_cut}
\Omega =\frac{1}{2\pi i}\int_{-\infty}^{\infty} d\omega
  \int \frac{d^3k}{(2\pi)^3}\,{\it disc} 
  \left[\log \chi(\omega+i\epsilon,k)\right] \, f_{BE}(\omega)\, , 
\ee
where $f_{BE}(\omega)=[\exp(\beta\omega)-1]^{-1}$ is the 
Bose-Einstein distribution. 

 In order to compute $\Omega$ to leading order in the fugacity 
$z=\exp(\beta\mu)$ it is sufficient to compute $\chi(\omega,k)$ to 
zeroth order in $z$. We get 
\be 
\label{chi_0}
 \chi(\omega,k) = \frac{m}{4\pi}\left\{ 
  im^{1/2}\left[ \omega-\frac{\epsilon_k}{2}+2\mu\right]^{1/2}
 -\frac{1}{a}\right\}.
\ee
It is now straightforward to compute the integral over $\omega$ 
and $k$ to leading order in $z$. On the BCS side $a<0$ we get 
\be 
\label{Omega_2}
\Omega = \frac{\sqrt{2}Tz^2}{\lambda^3}
 \exp\left( \beta B_2\right) 
 \left\{ 1 - {\it Erf} \left(\sqrt{\beta B_2}\right) \right\} \, , 
\ee
with $\beta=1/T$ and $B_2=1/(ma^2)$. On the BEC side there is 
an extra bound state contribution. This result can be compared to the 
virial expansion $\Omega=\nu z\lambda^{-3}(1+b_2 z+ O(z^2))$, where $\nu
=2$ is the number of degrees of freedom. The second virial coefficient 
$b_2=b_2^0+\delta b_2$ is the sum of free part $b_2^0=-1/(4\sqrt{2})$, which 
arises from quantum statistics, and an interacting contribution $\delta 
b_2$. Equ.~(\ref{Omega_2}) reproduces the standard result for the 
interaction contribution to the second virial coefficient
\be 
\label{b_2}
\delta b_2 = -\frac{{\it sgn}(a)}{\sqrt{2}}
  \exp(\beta B_2) \left\{ 1- {\it Erf} \left(\sqrt{\beta B_2}\right)
    \right\} + \Theta(a) \sqrt{2}  \exp(\beta B_2)\, . 
\ee
Near unitarity we have 
\be 
\label{b_2_a}
\delta b_2 = \frac{1}{\sqrt{2}} \left\{ 1+\frac{2}{\sqrt{\pi}}
  \frac{1}{\sqrt{mT}a} + \ldots \right\}\, .
\ee
Higher order terms in the fugacity expansion are discussed in 
\cite{Bedaque:2002xy,Liu:2009,Kaplan:2011br}.

\subsection{Fermion self energy}

\begin{figure}[t]
\bc\includegraphics[width=8cm]{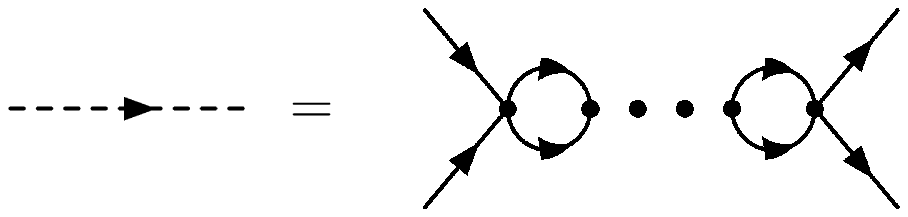}
\hspace*{2cm}
\includegraphics[width=4cm]{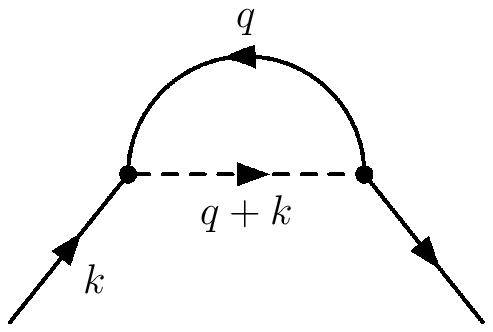}\ec
\vspace*{-3.5cm}\flushleft\hspace*{0cm} a)\hspace*{10cm}b)
\vspace*{2.5cm}
\caption{Boson propagator (Fig.~a) and fermion self energy (Fig.~b) 
in a dilute Fermi gas. Solid lines denote fermion propagators, the 
dashed line denotes the boson propagator.
\label{fig_sig}}
\end{figure}

 In order to determine the fermion self energy we leave 
the integration over $\psi$ and $\Phi$ in place, but 
write the lagrangian as the sum of a free and an interacting
term, ${\cal L}= {\cal L}_0 + {\cal L}_1$. In momentum space
\bea
\label{L_0}
{\cal L}_0 & =&  \psi^\dagger \left( \omega - \epsilon_p +\mu \right) \psi  
 + \Phi^*\chi_0(\omega,p)\Phi \, ,  \\
\label{L_1}
{\cal L}_1 & =& \left[ (\psi\sigma_+\psi)\Phi + {\it h.c.}\right]
 - \Phi^*\chi_0(\omega,p)\Phi\, , 
\eea
where $\chi_0(\omega,k)$ is the leading order polarization 
function given in equ.~(\ref{chi_0}). This lagrangian describes
an interacting theory of fermions of mass $m$ and bosons of 
mass $2m$. Note that the second term in equ.~(\ref{L_1}) 
renders the perturbative expansion well defined by 
removing fermion loop insertions in the boson propagator. 

The leading contribution to the fermion self energy come from 
the diagram shown in Fig.~\ref{fig_sig}. We find
\be
\Sigma(i\omega_n,k) = \frac{1}{2\pi i}\frac{4\pi}{m^{3/2}}
  \int d\Omega
 \int \frac{d^3q}{(2\pi)^3}
 \frac{f_{BE}(\Omega)}{\big[i\sqrt{\Omega+i\omega_n
               -\Xi_{q+k}}-(\sqrt{m}a)^{-1}\big]
           \big[\Omega-\xi_{q}\big]}\, , 
\ee
where the contour encircles the pole of the fermion propagator
and the branch cut of the boson propagator. We have defined 
$\xi_q=\epsilon_q-\mu$ with $\epsilon_q=q^2/(2m)$ and $\Xi_q=
\epsilon_q/2-2\mu$. The branch cut starts at ${\it Re}\,\Omega>
-2\mu$, and as a result the contribution from the cut is $O(z^2)$.
The leading $O(z)$ contribution to the self energy arises from the 
fermion pole. Analytically continuing this term to the on-shell 
point $i\omega_n=\xi_k$ we find
\be
\Sigma(k) = \frac{4\pi}{m^{3/2}}
 \int \frac{d^3q}{(2\pi)^3}
 \frac{f_{BE}(\xi_{q})}{i\sqrt{\xi_{q}+\xi_k-\Xi_{q+k}}-(\sqrt{m}a)^{-1}}\, . 
\ee
To leading order in $z$ we can replace the Bose-Einstein distribution
by a Boltzmann distribution. Near unitarity the real and imaginary parts 
of the on-shell self energy are 
\bea
\label{Re_Sig}
{\it Re}\,\Sigma(k) &=& -\frac{4\sqrt{2}zT}{\sqrt{\pi}} 
       \frac{1}{a\sqrt{mT}}\sqrt{\frac{T}{\epsilon_k}} 
      F_D\left(\sqrt{\frac{\epsilon_k}{T}}\right)\, , \\
\label{Im_Sig}
{\it Im}\,\Sigma(k) &=& -
         \frac{2zT}{\sqrt{\pi}} 
         \sqrt{\frac{T}{\epsilon_k}} 
    {\it Erf}\left(\sqrt{\frac{\epsilon_k}{T}}\right)\, .  
\eea
where $F_D$ is Dawson's Integral. In the high temperature 
limit we find fermion quasi-particles with energy $E_k=E_k^0
+\Delta E_k$ where $E_k^0=\epsilon_k$ and $\Delta E_k={\it 
Re}\,\Sigma(k)$. The width of the quasi-particles is 
given by $-{\it Im}\, \Sigma(k)$.

\section{Quasi-particle Boltzmann equation}
\label{sec_qp_boltz}
\subsection{Conservation laws}

 In kinetic theory the quasi-particles are described by a Boltzmann 
equation 
\be 
 \left( \frac{\partial}{\partial t} 
  + \left(\vec{\nabla}_pE_p\right) \cdot \vec{\nabla}_x
  - \left(\vec{\nabla}_xE_p\right) \cdot \vec{\nabla}_p \right)
f_p\left(\vec{x},t\right)
 = C[f_p]\, .
\ee
Here, $\vec{v}_p=\vec{\nabla}_pE_p$ is the quasi-particle velocity, 
$\vec{F}=-\vec{\nabla}_xE_p$ is the force term, and $C[f_p]$ is the 
collision term. The collision term conserves the number of particles
as well as their total momentum and energy.  This leads to three
three conserved currents. The conserved particle current is easy to 
find. Define
\bea
 n\left(\vec{x},t\right) &=& \int d\Gamma_p\,  f_p\left(\vec{x},t\right)\\
\vec\jmath (\vec{x},t) &=& \int d\Gamma_p\, \left(\vec{\nabla}_p E_p\right) 
         f_p\left(\vec{x},t\right)\, , 
\eea
where $d\Gamma_p=(d^3p)/(2\pi)^3$. Taking moments of the Boltzmann 
equation gives
\be 
 \frac{\partial}{\partial t} n\left(\vec{x},t\right)
 + \vec{\nabla}_x\cdot \vec\jmath (\vec{x},t) = 0 \, . 
\ee
Finding the correct form of the momentum conservation law is more 
complicated. We define the momentum density
\be 
\vec{\pi}\left(\vec{x},t\right) = \int d\Gamma_p\, \vec{p} 
  f_p \left(\vec{x},t\right)\, , 
\ee
and the  stress tensor
\be 
\label{pi_ij_kin}
\Pi^{ij}\left(\vec{x},t\right) = 
\int d\Gamma_p\, p^i\left(\vec{\nabla}^j_p E_p\right)  
   f_p\left(\vec{x},t\right)
 + \delta^{ij} \left( \int d\Gamma_p \, E_p f_p\left(\vec{x},t\right)
   - {\cal E}\left(\vec{x},t\right)\right)\,
\ee
where ${\cal E}$ is the energy density. Then 
\be 
 \frac{\partial}{\partial t} \pi^i\left(\vec{x},t\right)
 + \nabla^j_x \Pi^{ij} (\vec{x},t) = 0 \, ,
\ee
provided the quasi-particle energy $E_p$ and the energy density
${\cal E}$ satisfy the consistency condition
\be
\label{cons_cond}
 E_p = \frac{\delta {\cal E}}{\delta f_p}\, 
\ee
familiar from the theory of Fermi liquids. A second consistency 
condition arises from matching to hydrodynamics. In hydrodynamics
the momentum current is related to the particle current, $\vec{\pi}
=m\vec\jmath$. This implies
\be 
\label{cons_cond_2}
\int d\Gamma_p \, \vec{p}f_p\left(\vec{x},t\right)
=\int d\Gamma_p \, m\left(\vec{\nabla}_p E_p\right) 
   f_p\left(\vec{x},t\right)\, . 
\ee
This relation is automatically satisfied if $E_p$ can be 
written as 
\be 
E_p = E_p^0 + \int d\Gamma_{p'}\, T_{pp'} f_{p'}\, ,
\ee
as is the case for the results derived in the previous
section. Finally, the conserved energy current is
\be 
\vec{\jmath}_\epsilon \left(\vec{x},t\right)= \int d\Gamma_p\, E_p 
   \left(\vec{\nabla}_p E_p\right)   f_p\left(\vec{x},t\right)\, ,
\ee
which leads to 
\be
\frac{\partial}{\partial t} \, {\cal E} + 
  \vec{\nabla}\cdot \vec{\jmath}_\epsilon = 0 \, . 
\ee
In thermal equilibrium and in the local rest frame of the fluid 
the stress tensor is related to the pressure, $P=\frac{1}{3}
\Pi_{ii}$. Equ.~(\ref{pi_ij_kin}) then implies a simple formula
for the enthalpy, 
\be
{\cal E}+P=\int d\Gamma_p\, \left(
  \frac{1}{3}\vec{p}\cdot\vec{\nabla}_p E_p +E_p\right) f_p^0\, , 
\ee
where $f_p^0$ is the equilibrium distribution.

\subsection{Spin}

 So far we have ignored the role of spin. In a system with two 
spin degrees of freedom we have to sum the quasi-particle terms
over the spin degrees of freedom. The total density, for example, 
is given by
\be
  n\left(\vec{x},t\right) = \sum_{a=\uparrow,\downarrow} 
  \int d\Gamma_p\,  f^a_p\left(\vec{x},t\right)\, . 
\ee
The consistency condition for $E_p^a$ is 
\be 
 E^{a}_p = 
   \frac{\delta {\cal E}[f_p^\uparrow,f_p^\downarrow]}
        {\delta f^{a}_p}\,  .
\ee
At leading order in the fugacity the energy of an up-spin
is only a functional of the down-spin density (and vice 
versa). This implies
\be 
 \frac{\delta E^{\uparrow}_p}{\delta f_{p'}^\downarrow}
  = \frac{\delta E^{\downarrow}_{p}}{\delta f_{p'}^\uparrow}
  = T_{pp'}^{\uparrow\downarrow}\, ,
\hspace{0.5cm}
\frac{\delta E^{\uparrow}_p}{\delta f_{p'}^\uparrow}
  = \frac{\delta E^{\downarrow}_p}{\delta f_{p'}^\downarrow}
  = 0 \, . 
\ee

\subsection{Off-equilibrium bulk stress}

 In fluid dynamics the trace of the stress tensor in the rest frame 
of the fluid is given by $\Pi\equiv\frac{1}{3}\Pi_{ii} = P-\zeta
\left(\vec{\nabla}\cdot\vec{V}\right)$, where $\vec{V}$ is the fluid 
velocity. Computing the bulk viscosity requires calculating the 
dissipative part of the bulk stress $\Pi$. In kinetic theory we write
\be 
 \Pi[f_p^0+\delta f_p] \equiv  \Pi[f_p^0]+\delta\Pi
  \equiv \Pi^0 + \delta\Pi\, , 
\ee
where $f_p^0$ is the equilibrium distribution function and $\delta f_p$ 
is an off-equilibrium correction induced by the bulk flow $\left(
\vec{\nabla}\cdot\vec{V}\right)$. The distribution function is spin 
symmetric, $f_p\equiv f_p^\uparrow = f_p^\downarrow$. In order to compute 
$\delta\Pi$ we use equ.~(\ref{pi_ij_kin}), 
\be 
\Pi[f_p] = \frac{\nu}{3}
\int d\Gamma_p\, \left( \vec{p}\cdot \vec{\nabla} E_p \right) f_p
 +  \nu\int d\Gamma_p \, E_p f_p - {\cal E}[f_p]\, , 
\ee
and functionally expand ${\cal E}$ and $E_p$, 
\bea 
 {\cal E}[f^0_p+\delta f_p] &=& 
  {\cal E}^0 + \nu\int d\Gamma_p\, E_p\, \delta f_p 
     +  \frac{\nu}{2}\int d\Gamma_p\int d\Gamma_{p'}\, 
           T_{pp'} \delta f_p \delta f_{p'} +\ldots \, , \\
 E_p[f^0_p+\delta f_p] &=& 
     E_p + \int d\Gamma_{p'}\, T_{pp'} \delta f_{p'} 
 +\ldots\, , 
\eea
where we have used the consistency condition (\ref{cons_cond}).
In Sect.~\ref{sec_CE} we will determine $\delta f_p$ by solving 
the Boltzmann equation to leading order in the fugacity. We 
will find that $\delta f_p=O(z)$. At this order $\delta f_p$ 
satisfies the condition
\be 
\label{ortho_1}
 \int d\Gamma_p\, \delta f_p E_p^0 = 0\, ,
\ee
which has the simple interpretation that non-equilibrium effects
do not change the total energy of the fluid at $O(z)$. 
Equ.~(\ref{ortho_1}) implies that 
\be 
\label{ortho_2}
 \int d\Gamma_p\, \delta f_p E_p = 
       \int d\Gamma_p\, \delta f_p \Delta E_p\, .  
\ee
This result, together with $\delta f_p=O(z)$ and $\delta E_p=O(z)$,
means that the off-equilibrium bulk stress $\delta\Pi$ is of order 
$z^2$. In order to compute the dissipative part of the bulk stress 
to this accuracy we have to remove the $O(z^2)$ shift in the 
equilibrium pressure due to a shift in energy induced by $\delta f_p$. 
This corresponds to subtracting from $\delta\Pi$ the term $\delta P =
(\frac{\partial P}{\partial {\cal E}})\delta {\cal E}$ with $\frac{\partial P}
{\partial {\cal E}}=\frac{2}{3}$ and $\delta {\cal E}=\int d\Gamma_p\,
\delta f_p\Delta E_p$ (a similar subtraction in relativistic kinetic 
theory is discussed in \cite{Arnold:2006fz,Dusling:2011fd}). Putting 
all these ingredients together we find \footnote{The partial derivatives
with respect to $\mu,T$ arise from integration by parts together 
with $p\cdot\nabla_p f_p^0=-(2\mu\frac{\partial}{\partial\mu}+2T
\frac{\partial}{\partial T})f_p^0$, valid to leading order in $z$.}
\be 
\label{del_pi_fin}
\delta\Pi = \frac{\nu}{3}\int d\Gamma_p \, \delta f_p
  \left( \vec{p}\cdot\vec{\nabla}_p + 2\mu\frac{\partial}{\partial \mu}
    + 2T\frac{\partial}{\partial T} -2\right) \Delta E_p\, . 
\ee
This result has a simple interpretation as the shift in the pressure 
due to the scale breaking part of the quasi-particle energy. In 
particular, if $\Delta E_p$ has the scale invariant form $E_p 
\sim zTg(\epsilon_p/T)$ with an arbitrary function $g(x)$, then 
$\delta \Pi$ vanishes independently of the structure of $\delta f_p$.

\section{Solution of the Boltzmann equation}
\label{sec_CE}

 We determine $\delta f_p$ by solving the Boltzmann equation using
the standard Chapman-Enskog procedure. We write
\be 
 f_p(\vec{x},t) = f_p^0(\vec{x},t)
    \left( 1 - \frac{\psi_p}{T} \right)\, , \hspace{0.5cm}
 f_p^0(\vec{x},t) =
    \exp\left(-\frac{E(\vec{P},\vec{x},t)-\mu(\vec{x},t)}
  {T(\vec{x},t)}\right) \, , 
\ee
where $E(\vec{P},\vec{x},t)$ is the quasi-particle energy $E_p$ derived
in Sect.~\ref{sec_diag} evaluated for $\vec{P}=\vec{p}-m\vec{V}(\vec{x},t)$,
$\mu=\mu(\vec{x},t)$, and $T=T(\vec{x},t)$. At first order in the 
derivative expansion the off-equilibrium factor is 
\be 
\psi_p= \chi^B(\vec{p})\,\vec{\nabla}\cdot \vec{V}
       +\chi^S_{ij}(\vec{p})\sigma_{ij}
       +\chi^T_i(\vec{p}) \nabla_i T\, , 
\ee
with $\sigma_{ij}=\nabla_i V_j+\nabla_j V_i-\frac{2}{3}\delta_{ij}
\vec{\nabla}\cdot \vec{V}$. In this work we concentrate on the bulk
term $\chi_B$.

\subsection{Streaming term}
\label{sec_lhs}

 The left hand side of the Boltzmann equation is given by 
\be
{\cal D}f_p = \left( \frac{\partial}{\partial t} 
  + \vec{v}_p \cdot \vec{\nabla}_x
  + \vec{F} \cdot \vec{\nabla}_p\right) f_p\, . 
\ee
The streaming operator ${\cal D}$ acting on the equilibrium distribution 
function generates time derivatives and gradients of the thermodynamic 
variables $T,\mu$ and $\vec{V}$. We can write time derivatives in terms 
of spatial derivatives using the equations of fluid dynamics. At leading 
order in the derivative expansion it is sufficient to use ideal 
hydrodynamics. The equations can be further simplified by going to the 
local rest frame. We have 
\be 
\frac{\partial n}{\partial t} = -n\left(\vec{\nabla}\cdot\vec{V}\right)
\, , \hspace{0.5cm}
\frac{\partial s}{\partial t} = -s\left(\vec{\nabla}\cdot\vec{V}\right)
\, , \hspace{0.5cm}
\frac{\partial \vec{V}}{\partial t} = - \frac{1}{\rho}\vec{\nabla}P\,  ,
\ee
where $s$ is the entropy density, and $\rho$ is the mass density.
The first two equations imply that the entropy per particle is 
constant, $s/n={\it const}$. This means that the time derivatives
of all scalar thermodynamic variables can be expressed in terms 
of $\vec{\nabla}\cdot\vec{V}$. We find
\be 
\frac{\partial P}{\partial t} = -\rho c_s^2\left(\vec{\nabla}\cdot
  \vec{V}\right)
\, , \hspace{0.5cm}
\frac{\partial T}{\partial t} = -\frac{\rho\alpha T}{c_V} c_T^2
  \left(\vec{\nabla}\cdot\vec{V}\right)
\, , 
\ee
where $c_s$ and $c_T$ are the speed of sound at constant entropy per
particle and temperature, $c_V$ is the specific heat at constant volume, 
and $\alpha$ is the thermal expansion coefficient. See Appendix 
\ref{app_th} for definitions and explicit expression in terms of the 
equation of state. We can now collect all terms on the 
left hand side of the Boltzmann equation. We get 
\begin{eqnarray}
\frac{T}{f_0}{\cal D} f_0&=& \mbox{} \Bigg\{
  \frac{\alpha \rho c_T^2}{c_V} h - mc_s^2
  + \left[ \frac{1}{3}\vec{p}\cdot\vec{\nabla}_p 
             - \frac{\alpha \rho c_T^2}{c_V} 
  + \rho c_s^2 \left.\frac{\partial}{\partial P}\right|_T
             + \frac{\alpha \rho c_T^2}{c_V}T  
                      \left.\frac{\partial}{\partial T}\right|_P
\right] E_p\Bigg\}\, \vec{\nabla}\cdot \vec{V} \nonumber\\
&& \mbox{}\;
   + \left[\frac{E_p -h}{T}\vec{v}_p
     +\frac{c_P}{\alpha nT}\left(\vec{v}_p-\frac{\vec{p}}{m}\right)\right]
       \cdot \vec{\nabla} T
   + \frac{1}{2}\,  (v_p)_i p_j\sigma_{ij}\, , 
\label{lhs}
\end{eqnarray}
where $h$ is the enthalpy per particle. We observe that the term 
proportional to the bulk stress $\vec{\nabla}\cdot \vec{V}$ depends
in complicated ways on thermodynamic properties and the density 
and temperature dependence of $E_p$. There is an extra contribution
proportional to $(\vec{v}_p-\vec{p}/m)$ in the thermal conductivity
term, but because of the consistency condition (\ref{cons_cond_2}) 
this term vanishes when integrated against a distribution function.

 In the following we will focus on the term proportional to 
$(\vec{\nabla}\cdot \vec{V})$. This term can be simplified by
writing $E_p=\epsilon_p+\Delta E_P$ and dropping terms of 
order $z^2$. We get
\begin{eqnarray}
\left.\frac{T}{f_0}{\cal D} f_0\right|_{\it bulk}&=& \mbox{} \Bigg\{
  \frac{\alpha \rho c_T^2}{c_V} h - mc_s^2
  + \left[ \frac{2}{3} 
          - \frac{\alpha \rho c_T^2}{c_V} \right]\epsilon_p \nonumber \\
  & & \label{lhs_2} \mbox{} + \frac{1}{3}
   \left[\vec{p}\cdot\vec{\nabla}_p + 2\mu\frac{\partial}{\partial \mu}
    + 2T\frac{\partial}{\partial T} -2\right] \Delta E_p
      \Bigg\}\, \vec{\nabla}\cdot \vec{V} \, . 
\end{eqnarray}
We observe that the bulk viscosity term in the Boltzmann equation 
depends on the same scale breaking part of the quasi-particle energy 
that also appears in the bulk stress, equ.~(\ref{del_pi_fin}).
There a number of simple consistency checks for equ.~(\ref{lhs_2}). 
In a non-interacting gas $\Delta E_p=0$ and 
\be
h=\frac{5}{2}T,\hspace{0.3cm} 
c_P=\frac{5}{2}n,\hspace{0.3cm} 
c_V=\frac{3}{2}n,\hspace{0.3cm} 
c_s^2=\frac{5}{3}\frac{T}{m},\hspace{0.3cm} 
c_T^2=\frac{T}{m},\hspace{0.3cm} 
\alpha=\frac{1}{T}.
\ee
Using these values we find that the coefficient of the bulk stress 
vanishes. This result can be found in standard text books on kinetic theory
\cite{Landau:kin}. We also find that the bulk stress vanishes for a 
general scale invariant equation of state characterized by a 
temperature independent second virial coefficient, see Appendix 
\ref{app_th}. In order to compute the streaming term near unitarity
we use the second virial coefficient given in equ.~(\ref{b_2_a})
and the quasi-particle self energy in equ.~(\ref{Re_Sig}). We get
$\frac{T}{f_0}{\cal D} f_0 \equiv X_p (\vec{\nabla}\cdot\vec{V})$
with
\be
\label{X_p}
 X_p =  \frac{2\sqrt{2}}{9\sqrt{\pi}}
  \frac{zT}{a\sqrt{mT}}
  \left\{ \frac{\epsilon_p}{T} -\frac{9}{2}
     + 6\sqrt{\frac{T}{\epsilon_p}} 
        F_D\left(\sqrt{\frac{\epsilon_p}{T}}\right) \right\}\, . 
\ee
This result satisfies two non-trivial sum rules
\be 
\label{ortho_3}
 \int d\Gamma_p \, f^0_p X_p = 0\, , \hspace{0.5cm}
 \int d\Gamma_p \, f^0_p \epsilon_p X_p = 0 \, , 
\ee
which follow from the conservation of particle number and energy at 
leading order in $z$. Clearly, these sum rules can only be satisfied 
if the quasi-particle energy is consistent with the equation of state. 

\subsection{Collision term}
\label{sec_rhs}

 At leading order in the fugacity the collision term is dominated 
by two-body collisions. In the case of bulk stress the linearized 
collision operator is given by 
\be
 C[f^0_p+\delta f_p]\equiv \frac{f^0_p}{T} C_L[\chi_B(p)]
  \left(\vec{\nabla}\cdot\vec{V}\right)
\ee
with 
\be 
 C_L[\chi_B(p_1)] = \int \Big(\prod_{i=2}^4d\Gamma_{i}\Big)
   w(1,2;3,4) f^0_{p_2}
   \left[\chi_B(p_1)+\chi_B(p_2)-\chi_B(p_3)+\chi_B(p_4)\right]\, . 
\ee
The transition rate $w(1,2;3,4)$ is given by 
\be 
w(1,2;3,4) = (2\pi)^4\delta^3\Big(\sum_i \vec{p}_i\Big)
  \delta \Big( \sum_i E_{i}\Big) \left| {\cal A}\right|^2\, ,
\ee
and the scattering amplitude is 
\be 
 \left| {\cal A}\right|^2 = \frac{16\pi^2}{m^2} \frac{a^2}{a^2q^2+1}\, ,
\ee
where $\vec{q}=\frac{1}{2}\left(\vec{p}_2-\vec{p}_1\right)$. 
To leading order in $z$ we can approximate the quasi-particle
energy by the non-interacting result $E_p\simeq \epsilon_p$. 
Conservation of particle number and energy then leads to the 
sum rules given in equ.~(\ref{ortho_3}). In order to compute 
$\chi_B$ to leading order in $1/a$ we can also use the scattering 
amplitude in the unitary limit. We solve the linearized Boltzmann 
equation 
\be 
  X_p = C_L[\chi_B(p)]
\ee
by expanding $\chi_B(p)$ in generalized Laguerre (Sonine) polynomials
\be 
 \chi_B(p) = \sum_{i=2}^N c_i L^{1/2}_i
   \left(\frac{\epsilon_p}{T} \right) \, .
\ee
Restricting the sum to terms of order $i\geq 2$ ensures that the 
sum rules are satisfied. We solve for $c_i$ by taking moments
of the linearized Boltzmann equation, 
\be 
 \int d\Gamma_p\,  f^0_p L^{1/2}_k \Big(\frac{\epsilon_p}{T} \Big) X_p 
  =  \sum_i c_i \int d\Gamma_p\,  f^0_p L^{1/2}_k 
        \Big(\frac{\epsilon_p}{T} \Big)
        C_L\Big[ L^{1/2}_i \Big(\frac{\epsilon_p}{T} \Big) \Big]
\, , \hspace{0.5cm} (k=2,\ldots,N)\, . 
\ee
The simplest case is $N=2$. We find
\be 
 \chi_B(p) = \frac{\sqrt{\pi}}{64} 
   \frac{z}{a\sqrt{mT}}
  \left[ 15 -20\Big(\frac{\epsilon_p}{T} \Big)  
        + 4  \Big(\frac{\epsilon_p}{T} \Big)^2 \right]\, , 
\ee
and, using equ.~(\ref{del_pi_fin}), 
\be 
\label{zeta_fin}
\zeta=\frac{1}{24\sqrt{2}\pi}\lambda^{-3}
   \left(\frac{z\lambda}{a}\right)^2\, . 
\ee
We observe that $\chi_B$ is first order in the conformal breaking 
parameter $(\frac{z\lambda}{a})$ whereas the bulk viscosity is second 
order. The expansion in Laguerre polynomials converges rapidly. We 
can write $\zeta=k\lambda^{-3}(\frac{z\lambda}{a})^2$ where $k$ is 
a pure number. For $N=2,3,4$ we find 
\be 
k = \left\{ \frac{1}{24},
            \frac{9}{208},
            \frac{141461}{3258432} \right\}
                 \frac{1}{\sqrt{2}\pi},
\ee
corresponding to $k=\{9.378,9.739,9.771\}\cdot 10^{-3}$. The $N=3$ term
gives a 4\% correction, and the $N=4$ leads to a 0.3\% shift. 

\section{Outlook}
\label{sec_sum}

 The result in equ.~(\ref{zeta_fin}) can be written in the form 
\be 
\label{zeta_n}
\frac{\zeta}{n} = \frac{1}{9\sqrt{2\pi}} \frac{1}{(k_Fa)^2} 
  \left( \frac{T_F}{T}\right)^{5/2}\, 
\ee
where $k_F=(3\pi^2n)^{1/3}$ and $T_F=k_F^2/(2m)$ are defined in terms of the 
local density. We first address the question whether current experiments 
are sensitive to a bulk viscosity in this range. Measurements of the shear 
viscosity using collective modes are sensitive to values as small as 
$\eta/n\simeq 0.1$. The bulk viscosity grows with $1/(k_Fa)$ and $T_F/T$, 
so part of the issue is how far one can extrapolate our result in these 
two variables. We know that the bulk viscosity vanishes for both $|k_Fa|
\to \infty$ and $|k_Fa|\to 0$. This means that at fixed $T/T_F$ the bulk 
viscosity has a maximum at some finite value of $(k_Fa)$. Independent 
of the location of this maximum we also know that in typical experiments 
hydrodynamics breaks down for $|k_Fa|\gsim 1$ \cite{Kinast:2004}. As a 
function of $T/T_F$ we expect the bulk viscosity to have a maximum near 
the phase transition, $T\sim T_c \simeq 0.167(13)T_F$ \cite{Ku:2012}. 
In the case of shear viscosity we know that kinetic theory is remarkably 
accurate down to temperatures as low as $T\sim 2T_c$, see for example 
\cite{Enss:2010qh}. Using $|k_Fa|\sim 1$ and $T\sim 2T_c$ in 
equ.~(\ref{zeta_n}) we conclude that $\zeta/n$ could be as large as 0.5, 
within the range accessible in experiment. 

 Additional information on the temperature dependence of $\zeta$ 
is provided by calculations in the low temperature, superfluid, phase. 
A superfluid is characterized by three bulk viscosity coefficients, 
$\zeta_1,\zeta_2$ and $\zeta_3$. Of these, $\zeta_2$ is the coefficient
that is analogous to the bulk viscosity in the normal phase. Using a 
kinetic theory of phonons Escobedo et al.~find \cite{Escobedo:2009bh}
\be 
\zeta_2 = \frac{19\pi^4}{2} d_0^2 c_2^2\xi^{9/2} 
  \frac{1}{a^2m\mu} \frac{m^{3/2}T^3}{\mu^{3/2}}\, . 
\ee
Here, $d_0$, $c_2$, and $\xi$ are non-perturbative coefficients related
to the equation of state and the phonon dispersion relation. The Bertsch
parameter $\xi\simeq 0.38$ \cite{Forbes:2010gt} governs the relationship 
between the chemical potential and the Fermi energy. The quantity $d_0$ 
is related to the $T=0$ value of the contact density. Using the results in 
\cite{Braaten:2010if} we estimate $d_0\simeq 0.3$. The parameter $c_2$ 
is one of two coefficients $c_{1,2}$ that govern the phonon dispersion 
relation. In principle $c_2$ can be extracted from the transverse current 
response at low momentum. Using a calculation at next-to-leading order in 
the epsilon expansion we found $c_1\simeq -0.020$ \cite{Rupak:2008xq}. At 
this order $c_2$ vanishes, and we will assume $|c_2|\lsim |c_1|$
\footnote{The mean field calculation in \cite{Manes:2008} gives 
$c_2=-0.003$. This value is indeed smaller than $c_1$, but it violates the 
constraint $c_2>0$ \cite{Son:2005rv}.}. Putting all these estimates
together we find
\be 
\frac{\zeta_2}{n}\lsim 0.1 \frac{1}{(k_Fa)^2}
                       \left(\frac{T}{T_F}\right)^3\, .
\ee 
The scaling is consistent with equ.~(\ref{zeta_n}) and the idea
that $\zeta$ has a maximum as a function of $T/T_F$. We note, however,
that $\zeta_2/n$ is very small for $T\lsim T_c$, suggesting that the 
maximum occurs in the normal phase. 

 Another interesting issue concerns the frequency dependence of 
the bulk viscosity. Taylor and Randeria proved the sum rule
\cite{Taylor:2010ju,Goldberger:2011hh}
\be 
 \frac{1}{\pi} \int d\omega\, \zeta(\omega) = 
  \frac{1}{72\pi ma^2} 
  \left. \frac{\partial {\cal C}}{\partial a^{-1}}\right|_{s/n}\, .
\ee
Using the virial expansion we find 
\be 
 \frac{1}{\pi} \int d\omega\, \zeta(\omega) \sim 
  T \lambda^{-3}\left(\frac{z\lambda}{a}\right)^2\, . 
\ee
We conclude that the kinetic theory result is consistent with this 
sum rule provided the width of the transport peak is less than $T$. 
Note that the width of the shear peak is $\tau_R^{-1}=P/\eta\sim 
zT\ll T$ \cite{Braby:2010tk}. The high frequency tail of the 
bulk viscosity was determined in \cite{Hofmann:2011qs},
\be 
 \zeta(\omega) = \frac{{\cal C}}{36\pi\sqrt{m\omega}}
  \frac{1}{1+a^2m\omega}\, .
\ee
In the high temperature limit this implies
\be 
\zeta(\omega) \sim \lambda^{-3}\left(\frac{z\lambda}{a}\right)^2
  \left( \frac{T}{\omega}\right)^{3/2}\, , 
\ee
showing that the transport peak and the high frequency tail 
can match smoothly if the transport peak is broad, with a 
width of order $T$. On the other hand, if the transport peak is 
narrow then the sum rule is saturated by the continuum contribution,
and the spectral function must have two peaks, a transport peak at 
$\omega=0$ and a continuum peak at $\omega\sim T$. 

 There are a number of issues that remain to be addressed. Based
on the discussion above it would be interesting to compute the 
frequency dependence of the bulk viscosity in kinetic theory. 
It would also be interesting to generalize our calculation to 
a two-dimensional Fermi gas. In two dimensions scale invariance 
is always broken but experiments indicate that the bulk viscosity 
is very small \cite{Vogt:2011np,Taylor:2012pe}. Finally, it 
would be interesting to construct a complete quasi-particle
model of the dilute Fermi gas near unitarity. We have found 
that the leading shift in the energy density due to scale breaking 
effects is completely determined by the real part of the self
energy, but at unitarity the self energy is imaginary
and the interaction part of the energy density cannot be written 
in terms of ${\it Re}\Sigma$ alone.

 Acknowledgments: We thank Paulo Bedaque and John Thomas for useful 
discussions, and C.~Manuel and Juan Torres-Rincon for comments on 
the manuscript. This work was supported in parts by the US Department 
of Energy grant DE-FG02-03ER41260. 

\begin{appendix}
\section{Thermodynamics}
\label{app_th}
\subsection{General relations}
\label{app_gen}

 Consider an equation of state in the form $P=P(\mu,T)$. Derivatives 
of the pressure with respect to $\mu$ and $T$ determine the entropy 
density and pressure
\begin{equation}
\label{s_and_n}
s=\left.\frac{\partial P}{\partial T}\right|_{\mu}\, , \hspace{1cm}
n=\left.\frac{\partial P}{\partial \mu}\right|_{T}\, . 
\end{equation}
The energy density is determined by the relation
\be
{\cal E}=\mu n +sT -P\, , 
\ee
and the enthalpy per particle is $h=({\cal E}+P)/n$. In order to compute 
the specific heat at constant volume we use $V=N/n$ and write 
\be
\label{c_v_part}
 c_{V}= \frac{T}{V}\left.\frac{\partial S}{\partial T}\right|_{V}
      = \frac{\partial(s,V)}{\partial(T,V)}
      = \frac{\partial(s,V)/\partial(T,\mu)}{\partial(T,V)/\partial(T,\mu)}
      = T\left[\left.\frac{\partial s}{\partial T}\right|_{\mu}
          -\frac{[(\partial n/\partial T)|_{\mu}]^2}
                {(\partial n/\partial \mu)|_{T}}\right]\, ,
\ee
where we have defined the Jacobian 
\be 
\frac{\partial(u,v)}{\partial(x,y)} = \left|
\begin{array}{ll}
 \frac{\partial u}{\partial x} &  \frac{\partial u}{\partial y} \\
 \frac{\partial v}{\partial x} &  \frac{\partial v}{\partial y} 
\end{array}\right|\, . 
\ee
In order to compute $c_P$ we make use of the relation between 
$c_P-c_V$ and the thermal expansion coefficient $\alpha=(1/V)(\partial V/
\partial T)|_P$. This relation is given by 
\begin{equation}
\label{cP}
   c_{P}-c_{V}=-\frac{T}{V}\frac{[(\partial V/\partial T)|_{P}]^{2}}
                      {(\partial V/\partial P)|_{T}}\, . 
\end{equation}
The partial derivatives are 
\begin{equation}
  \left. \frac{1}{V}\frac{\partial V}{\partial T}\right|_{P}
   =\frac{1}{n}\left[\frac{s}{n}
             \left.\frac{\partial n}{\partial \mu} \right|_{T}
      -\left.\frac{\partial n}{\partial T}\right|_{\mu}\right] \, , 
\hspace{1cm}
   \left. \frac{1}{V}\frac{\partial V}{\partial P}\right|_{T}
      =-\left. \frac{1}{n^2}\frac{\partial n}{\partial \mu}\right|_{T}\, , 
\end{equation}
which gives
\begin{equation}\label{}
    c_{P}=c_{V}+T\frac{\Big[
    \frac{s}{n} (\partial n/\partial \mu)\big|_{T}
     -(\partial n/\partial T)\big|_{\mu}\Big]^{2}}
              {(\partial n/\partial \mu)\big|_{T}}\, . 
\end{equation}
The isothermal and the adiabatic speed of sound are defined by 
\be 
 c_T^2 = \left.\frac{\partial P}{\partial \rho}\right|_T,
\hspace{1cm}
 c_s^2 = \left.\frac{\partial P}{\partial \rho}\right|_{s/n}.
\ee
We have 
\be
 c_T^2 = \frac{n}{m} \left[\left.\frac{\partial n}{\partial \mu}
\right|_T\right]^{-1}\, ,
\hspace{1cm}
 c_s^2 = \frac{c_p}{c_V}c_T^2\, ,
\ee
and the thermal expansion coefficient can be written as
\be 
\alpha = \frac{1}{T}\left[\frac{1}{c_T^2}\frac{T}{m} 
   \frac{c_P-c_V}{n}\right]^{1/2}\, .
\ee

\subsection{Virial Expansion}
\label{app_vir}

 We can use these results to determine thermodynamic properties 
from the virial expansion. At second order we have $P=\frac{\nu T}
{\lambda^{3}}(z+b_{2}z^2)$. At unitarity $b_2$ is a constant, but 
in general $b_2=b_2(T)$. The temperature dependence of $b_2$ is a 
measure of scale breaking. We find
\be
\frac{P-\frac{2}{3}{\cal E}}{P} = -\frac{2}{3}z Tb_2'(T)\, .
\ee
The enthalpy per particle is 
\be 
h=\frac{5T}{2}\left[1-z\left(b_2-\frac{2}{5}Tb_2^\prime\right)\right]\, 
\ee
and the specific heats are given by
\bea
  c_{V} &=& \frac{\nu z}{\lambda^{3}}\left[\frac{3}{2}
      + \frac{15}{4}zb_{2}(T)-zT b^\prime_{2}(T)
      + zT^{2} b^{\prime\prime}_{2}(T)\right]\, ,  
      \\ 
  c_{P} &=& c_{V}+\frac{\nu z}{\lambda^{3}}\Big[1+5zb_{2}(T)
            - 2zT b^\prime_{2}(T)\Big]\, . 
\eea
Finally, the speed of sound at constant $T$ and $s/n$ as well 
as the thermal expansion coefficient are 
\bea
c_T^2 &=& \;\frac{T}{m}\;\, \Big[ 1 -2zb_2(T)\Big]\, , \\
c_s^2 &=& \frac{5T}{3m}  \left[ 1 - z
           \left( b_2(T)+\frac{8}{15}T b_2^\prime(T)
                        +\frac{4}{15}T^2 b_2^{\prime\prime}(T)
                                            \right)\right]\, ,\\
\alpha &=& \;\frac{1}{T}\;  \left[ 1 + z
           \left(\frac{5}{2}b_2(T) - T b_2^\prime(T)\right)\right]\, . 
\eea
\end{appendix}


\end{document}